\documentclass[aps,reprint,amsmath,amssymb,superscriptaddress]
{revtex4-1}
\usepackage{graphicx}
\usepackage{dcolumn}
\usepackage{bm}
\usepackage{color} 
\usepackage{physics}
\usepackage{hyperref}
\usepackage{subfigure}
\hypersetup{
    colorlinks = true,
    linkcolor = blue,
    citecolor = blue,
    anchorcolor = blue,
    urlcolor = blue
    }

\def\lesssim{\ \raise.3ex\hbox{$<$}\kern-0.8em\lower.7ex\hbox{$\sim$}\ }
\def\gesim{\ \raise.3ex\hbox{$>$}\kern-0.8em\lower.7ex\hbox{$\sim$}\ }
\def\up{\uparrow}
\def\dwn{\downarrow}

\newcommand \beq{\begin{eqnarray}}
\newcommand \eeq{\end{eqnarray}}

\usepackage{amsmath,amssymb}
\usepackage{fixmath}
\usepackage[normalem]{ulem}

\usepackage{comment}

\usepackage{xcolor}
\definecolor{honey}{HTML}{ec9706}

\begin{document}

\title{Spin transport between polarized Fermi gases near the ferromagnetic phase transition}

\author{Tingyu Zhang}
\affiliation{Department of Physics, School of Science, The University of Tokyo, Tokyo 113-0033, Japan}
\affiliation{Interdisciplinary Theoretical and Mathematical Sciences Program (iTHEMS), RIKEN, Wako, Saitama 351-0198, Japan}
\author{Daigo Oue}
\affiliation{Instituto de Telecomunica\c{c}\~{o}es, Instituto Superior T\'{e}cnico, University of Lisbon, 1049-001 Lisbon, Portugal}
\affiliation{The Blackett Laboratory, Department of Physics, Imperial College London, Prince Consort Road, Kensington, London SW7 2AZ, United Kingdom}
\affiliation{
Kavli Institute for Theoretical Sciences, University of Chinese Academy of Sciences, Beijing, 100190, China.
}
\author{Hiroyuki Tajima}
\affiliation{Department of Physics, School of Science, The University of Tokyo, Tokyo 113-0033, Japan}
\author{Mamoru Matsuo}
\affiliation{
Kavli Institute for Theoretical Sciences, University of Chinese Academy of Sciences, Beijing, 100190, China.
}
\affiliation{CAS Center for Excellence in Topological Quantum Computation, University of Chinese Academy of Sciences, Beijing 100190, China}
\affiliation{Advanced Science Research Center, Japan Atomic Energy Agency, Tokai, 319-1195, Japan}
\affiliation{RIKEN Center for Emergent Matter Science (CEMS), Wako, Saitama 351-0198, Japan}
\author{Haozhao Liang}
\affiliation{Department of Physics, School of Science, The University of Tokyo, Tokyo 113-0033, Japan}
\affiliation{Interdisciplinary Theoretical and Mathematical Sciences Program (iTHEMS), RIKEN, Wako, Saitama 351-0198, Japan}

\begin{abstract}
We theoretically study the spin current between two polarized Fermi gases with repulsive interactions near the itinerant ferromagnetic phase transition.
We consider a two-terminal model where the left reservoir is fixed to be fully polarized while the polarization of the right reservoir is tuned through a fictitious magnetic field defined by the chemical-potential difference between different atomic hyperfine states.
We calculate the spectra of the spin-flip susceptibility function, which displays a magnon dispersion emerging from the Stoner continuum at low momentum in the ferromagnetic phase.
Based on the spin-flip susceptibility and using Keldysh Green's function formalism, we investigate the spin current induced by quasiparticle and spin-flip tunneling processes, respectively, and show their dependence on the polarization bias between two reservoirs.
The one-body (quasiparticle) tunneling demonstrates a linear dependence with respect to the polarization bias.
In contrast, the spin-flip process manifests a predominantly cubic dependence on the bias.
While indicating an enhanced magnon tunneling in the strong-coupling regime, our results also demonstrate a characteristic behavior around the critical repulsive strength for ferromagnetic phase transition at low temperatures.  
\end{abstract}

\maketitle

\section{Introduction}

The study of transport phenomena allows us to deeply understand the physical properties of various quantum many-body systems and has attracted a lot of attention in the research of cold atomic systems~\cite{chien2015quantum,Krinner_2017}.
In spin-balanced ultracold fermion systems, controlling Feshbach resonances~\cite{RevModPhys.82.1225} enables us to tune the scattering (and hence the interaction) between atoms; transport phenomena can be investigated in terms of the crossover between the Bardeen-Cooper-Schrieffer (BCS) and Bose-Einstein condensation (BEC) regimes, with an attractive interaction increasing monotonically~\cite{PhysRevLett.92.040403,PhysRevLett.92.120401,BCS-BEC}.
Owing to the controllability, experiments with such ultracold Fermi gases have been carried out in various regimes to observe, e.g., multiple Andreev reflections~\cite{doi:10.1126/science.aac9584}, and the AC and DC Josephson currents~\cite{science350,Krinner_2017,doi:10.1146/annurev-conmatphys-031218-013732,science369}.

On the repulsive side of the Feshbach resonance, a two-component Fermi gas undergoes a ferromagnetic phase transition below the Curie temperature as predicted by the mean-field Stoner model~\cite{Stoner}.
The magnetic moments of particles (corresponding to the hyperfine states called pseudospin in Fermi atomic gases) tend to be aligned parallel to each other under the effect of a sufficiently strong repulsion, giving rise to a transition from the paramagnetic to ferromagnetic states.
Experiments have provided strong evidence for the ferromagnetic phase transition, when the interaction strength exceeds a certain critical value~\cite{science.1177112}.
More recently, by investigating the spin dynamics of an ultracold $^6{\rm Li}$ gas, a scattering length $a$ for the critical interaction strength is found to be $k_{\rm F}a\simeq1$ at the temperature $T/T_{\rm F}\simeq0.12$~\cite{valtolina2017}, where $k_{\rm F}$ ($T_{\rm F}$) is the Fermi momentum (temperature).
On the other hand, a beyond mean-field theory has predicted a critical interaction strength around $k_{\rm F}a= 1.05$~\cite{PhysRevLett.95.230403,PhysRevA.79.053606} for a phase transition at low temperature, while quantum Monte-Carlo calculations give a lower value ($k_{\rm F}a\simeq 0.8$)~\cite{PhysRevLett.103.207201,PhysRevLett.105.030405} for zero temperature.
Variational calculations for the Fermi gas with the hard-sphere-potential approximation have shifted the transition to a higher repulsion strength as $k_{\rm F}a\simeq1.8$~\cite{PhysRevA.85.033615}. 
 
Apart from the quantitative analysis of the ferromagnetic transition beyond the mean-field theory, spin transport phenomena in spin-imbalanced systems have gained both experimental and theoretical attention in various systems~\cite{PhysRevLett.90.166602,PhysRevB.96.134412,PhysRevB.99.144411,linder2015,PhysRevLett.118.105303,han2020spin,PhysRevB.102.144521}.
In solid-state physics, particularly within the spintronics community, the spin tunneling in ferromagnet has been extensively discussed based on the spin Seebeck effect induced by a temperature gradient~\cite{uchida2008observation,jaworski2010observation,uchida2010spin,PhysRevB.81.214418,PhysRevB.83.094410,Adachi_2013,PhysRevLett.120.037201} and the spin pumping protocol realized by ferromagnetic resonance under microwave irradiation~\cite{PhysRevLett.88.117601,PhysRevLett.90.166602,10.1063/1.2199473,kajiwara2010transmission,PhysRevB.89.174417}.
In cold atomic systems, the bulk spin transport has been investigated theoretically in Fermi gases~\cite{PhysRevA.88.033630,PhysRevResearch.4.043014}, and has been experimentally explored during the transverse demagnetization process of a 3D Fermi gas~\cite{science.1247425}. Meanwhile, the mesoscopic spin transport phenomena at the interface have also been widely studied via Hamiltonian approach with a two-terminal model~\cite{doi:10.1073/pnas.1601812113}, where two polarized Fermi gases are connected through a quantum point contact~\cite{PhysRevA.105.043313}.
For normal Fermi gases, spin currents can be induced by the spin imbalance between two reservoirs with different polarization~\cite{PhysRevResearch.2.023152}.
This observation serves as motivation to infer that spin tunneling between Fermi gas clouds may exist due to the interaction.
\begin{figure}[t]
    \centering
    \includegraphics[width=8.6cm]{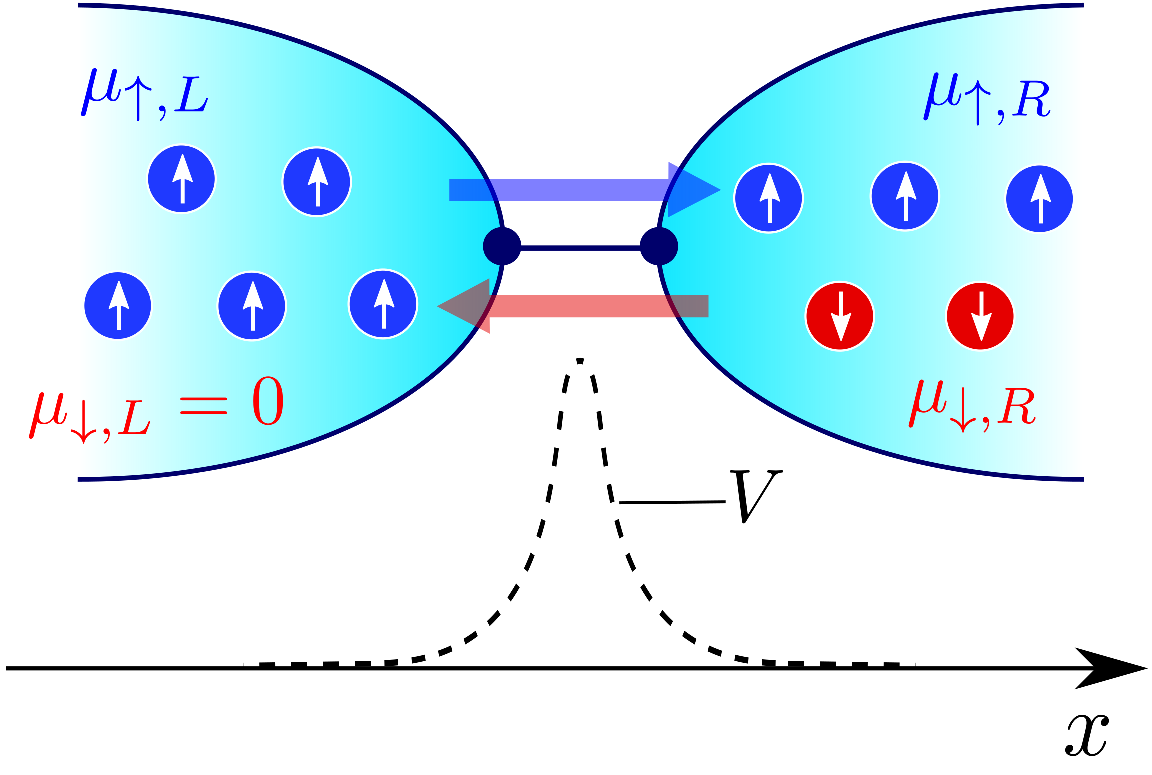}
    \caption{Schematic view of the two-terminal system considered in this work.
    The left reservoir is fully polarized while the right one is partially polarized. Spin currents are induced through the potential barrier due to the spin imbalance between two reservoirs. The external potential of the junction is shown below the schematic. The $\mu_{\sigma,i}$ represents the chemical potential for each component in each reservoir.
    }
    \label{schematic}
\end{figure}

Toward the cold-atomic quantum simulation of transport phenomena associated with spintronics, the multi-particle tunneling process, such as spin-flip tunneling at the interface, plays a crucial role, where the spin is exchanged between reservoirs.
However, the existence of such a process for cold atomic junctions is still elusive, because it is not straightforward to distinguish it from the usual quasiparticle tunneling.
On the other hand, the recent experimental results of anomalous tunneling transports in strongly interacting Fermi gases imply the possible existence of pair-tunneling transport in atomic systems.
Such a correlated tunneling event has attracted much interest in nuclear systems, where the two-nucleon pairing can suppress the sequential tunneling of nucleons~\cite{PhysRevLett.89.102501}. 
In Ref.~\cite{10.1093/pnasnexus/pgad045}, it is proposed that the noise measurement can be direct evidence of pair-tunneling transport.
Accordingly, it is an interesting question whether or not the spin-flip tunneling process can be identified via the spin-transport measurement in cold atomic systems.

In this work, we propose a system provoking the spin-flip tunneling through the potential barrier between two reservoirs consisting of two-component Fermi gases with the spin polarization (see Fig.~\ref{schematic}).
Note that an internal magnetic field is induced due to the spin polarization (i.e., finite chemical potential difference between two spin species $h=(\mu_{\up}-\mu_{\dwn})/2$, where $\mu_{\sigma}$ is the chemical potential for the spin $\sigma=\uparrow,\downarrow$). We focus on a regime with repulsive interactions.
The ferromagnetic phase corresponds to the fully spin-polarized regime induced by the effective magnetic field $h$ (explicit breaking of the spin-inversion symmetry) or the repulsive interaction leading to the Stoner instability.
 The reservoir on the left side is fixed to be fully polarized, where the minority chemical potential is sufficiently small (e.g., $\mu_{\dwn}=0$ at $T=0$). 
The magnetization of the right reservoir is adjusted to match that of the left reservoir, resulting in a state without spin bias. 
Alternatively, the magnetization of the right reservoir can be configured such that the majority spin direction is opposite to that of the left reservoir, leading to a substantial spin bias.
Such a situation is similar to the junction system where the left reservoir can be regarded as an analog of a half-metallic ferromagnet~\cite{PhysRevB.94.184405} and the right one may exhibit metallic or itinerant ferromagnetic behavior depending on the tuning parameters $h$ and $a$.
The average Fermi energy of two spin components $\epsilon _{\rm F}$ for the two reservoirs are set to be equal so that we can purely study the spin current without the mass tunneling current through the junction.

This paper is organized as follows. In Sec.~\ref{Hamiltonian}, we present the formalism: a two-terminal model and spin-flip tunneling current operators with a mean-field approximation. In Sec.~\ref{One-body}, we derive the formula of a one-body spin tunneling current up to the leading order and show its dependence on the spin bias between two reservoirs. In Sec.~\ref{spin-flip}, we adopt the random-phase approximation (RPA) to investigate the spin-flip susceptibility and numerically evaluate the spin-flip tunneling current. We conclude this work in Sec.~\ref{conclusion}.

Throughout the paper, we take $\hbar=k_B=1$ and the volumes for both reservoirs to be unity.

\section{Hamiltonian}\label{Hamiltonian}

The Total Hamiltonian of the two-terminal model for two-components Fermi gases 
with contact-type repulsive interaction is given by $H=H_{\rm L}+H_{\rm R}+H_{\rm T}$ (see Appendix~\ref{appendixA}), where the reservoir Hamiltonian $H_{i={\rm L},{\rm R}}$ reads
\begin{align}\label{Hreservoir}
    &H_{i={\rm L},{\rm R}}= \sum_{\bm{p},\sigma}\varepsilon_{\bm{p},\sigma,i}c^\dagger_{\bm{p},\sigma,i}c_{\bm{p},\sigma,i}\nonumber\\ &+g\sum_{\bm{p},\bm{p}',\bm{k}}c^\dagger_{\bm{p}+\frac{\bm{k}}{2},\uparrow,i}c^\dagger_{-\bm{p}+\frac{\bm{k}}{2},\downarrow,i}c_{-\bm{p}'+\frac{\bm{k}}{2},\downarrow,i}c_{\bm{p}'+\frac{\bm{k}}{2},\uparrow,i}.
\end{align}
Here $\varepsilon_{\bm{p},\sigma,i}=p^2/(2m)-\mu_{\sigma,i}$ is the kinetic energy of a Fermi atom with mass $m$ in the reservoir $i={\rm L,R}$ measured from the reservoir chemical potential $\mu_{\sigma,i}$, $g$ is the interaction strength, and $c^\dagger_{\bm{p},\sigma,i}$ ($c_{\bm{p},\sigma,i}$) creates (annihilates) a particle with spin $\sigma$ and momentum $\bm{p}$ in the reservoir $i$.
Also we can obtain the tunneling Hamiltonian eventually leading to the spin tunneling from one reservoir to the other as
\begin{subequations}\label{Htunneling}
    \begin{align}
        H_{\rm T}&=H_{\rm 1T}+H_{\rm 2T},\label{HTa}\\
        H_{\rm 1T}&= \mathcal{T}_1\sum_{\bm{p},\bm{q},\sigma}c^\dagger_{\bm{q},\sigma,{\rm R}}c_{\bm{p},\sigma,{\rm L}}+\rm{H.c.},\label{HTb}\\
        H_{\rm 2T}&=\mathcal{T}_2\sum_{\bm{p},\bm{q}}\big(S^+_{\bm{p},{\rm L}}S^-_{\bm{q},{\rm R}}+S^+_{\bm{q},{\rm R}}S^-_{\bm{p},{\rm L}}\big),\label{HTc}
    \end{align}
\end{subequations}
where we have kept the leading-order terms with respect to the single-particle transmission and reflection amplitudes~\cite{PhysRevA.106.033310} and introduced spin ladder operators, $S^+_{\bm{p},i}=c^\dagger_{\bm{p},\uparrow,i}c_{\bm{p},\downarrow,i}$ and $S^-_{\bm{p},i}=c^\dagger_{\bm{p},\downarrow,i}c_{\bm{p},\uparrow,i}$. Notice that $H_{\rm 1T}$ denotes the tunneling of a single particle with spin $\sigma$, while $H_{\rm 2T}$ denotes the exchange of the spin degrees of freedom between the left and right reservoirs. 
Therefore, $H_{\rm 1T}$ ($H_{\rm 2T}$) represents the one-body (spin-flip) tunneling process with the tunneling strength $\mathcal{T}_1$ ($\mathcal{T}_2$), which can be estimated with the one-particle transmission coefficient $B_{\bm{p},\sigma}$.
Assuming the long-wavelength limit for transmitted waves, we can write $\mathcal{T}_2=2g\operatorname{Re}[B^*_{\bm{0},\up}B_{\bm{0},\dwn}]$~\cite{PhysRevA.106.033310,10.1093/pnasnexus/pgad045}. 
Although $\mathcal{T}_1$ should generally depend on the spin component, we will estimate its averaged value using an average Fermi energy in each reservoir.
Moreover, while the low-energy effective tunneling process in Refs.~\cite{PhysRevA.106.033310,10.1093/pnasnexus/pgad045}
may conserve the momentum at the interface as in the case of uniform RF transport~\cite{zhang2023dominant},
we employ the momentum-unconserved tunneling~\cite{PhysRevResearch.2.023152}.
This simplification would not change the results qualitatively.

The spin-current operator in the Heisenberg representation is defined as 
\begin{align}
    \hat{I}_{\rm s}=&-\dot{N}_{\uparrow,{\rm L}}+\dot{N}_{\downarrow,{\rm L}}\nonumber\\
    =&~i[N_{\uparrow,{\rm L}},H_{\rm T}]-i[N_{\downarrow,{\rm L}},H_{\rm T}],
\end{align}
where $N_{\sigma,i}=\sum_{\bm{p}}c^\dagger_{\bm{p},\sigma,i}c_{\bm{p},\sigma,i}$ is the particle-number operator for each component. The current includes both contributions from the one-body and spin-flip tunneling processes, $\hat{I}_{\rm s}=\hat{I}_{\rm 1s}+\hat{I}_{\rm 2s}$, where we defined the one-body and spin-flip tunneling currents as
\begin{subequations}
    \begin{align}
        \hat{I}_{\rm 1s} =&~i[N_{\up,{\rm L}},H_{\rm 1T}]-i[N_{\dwn,{\rm L}},H_{\rm 1T}]\notag\\
        = &-i\mathcal{T}_1\sum_{\bm{p},\bm{q}}(c^\dagger_{\bm{q},\up,{\rm R}}
        c_{\bm{p},\up,{\rm L}}-c^\dagger_{\bm{q},\dwn,{\rm R}}
        c_{\bm{p},\dwn,{\rm L}})+{\rm H.c.},\label{I1s}\\
        \hat{I}_{\rm 2s}=&~i[N_{\up,{\rm L}},H_{\rm 2T}]-i[N_{\dwn,{\rm L}},H_{\rm 2T}]\notag\\
        =&~2i\mathcal{T}_2\sum_{\bm{p},\bm{q}}(S^\dagger_{\bm{p},{\rm L}}S^-_{\bm{q},{\rm R}}-S^+_{\bm{q},{\rm R}}S^-_{\bm{p},{\rm L}}).\label{I2s}
    \end{align}
\end{subequations}
Notice that the quantum tunneling with a similar two-terminal model has been studied for a strongly-correlated Fermi system~\cite{PhysRevA.106.033310}, which can be applied to nuclear reaction and magnonic spin transport.

For the evaluation of the expectation value of the tunneling currents in the following sections, we employ the mean-field approximation for the reservoirs.
The mean-field theory yields an effective chemical potential for each reservoir, $\mu'_{\sigma,i}=\mu_{\sigma,i}-gn_{\bar{\sigma},i}$, and a reservoir Hamiltonian 
\begin{equation}
    H_i=\sum_{\bm{p},\sigma}\xi^\sigma_{\bm{p},i}c^\dagger_{\bm{p},\sigma,i}c_{\bm{p},\sigma,i}-gn_{\up,i}n_{\dwn,i},
\end{equation}
where $n_{\sigma,i}=\sum_{\bm{p}}\langle c^\dagger_{\bm{p},\sigma,i}c_{\bm{p},\sigma,i}\rangle$ is the particle number density for each component under the mean-field approximation, and we defined $\xi^\sigma_{\bm{p},i}=\bm{p}^2/2m-\mu'_{\sigma,i}$.

\begin{figure}[t]
    \centering
    \includegraphics[width=8.6cm]{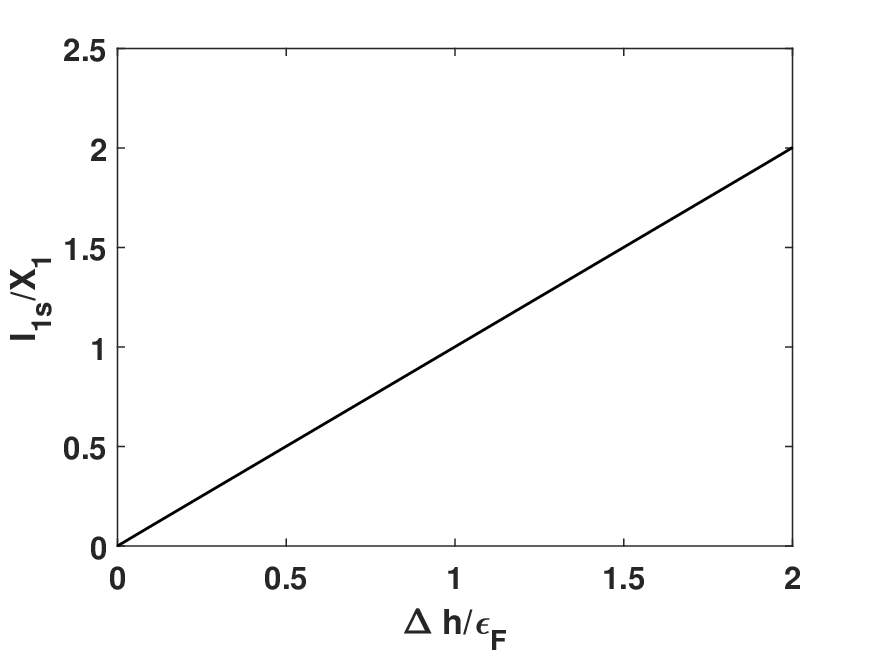}
    \caption{The one-body spin tunneling current as a function of polarization 
    bias $\Delta h=h_{\rm L}-h _{\rm R}$, where $X_1$ is the normalizing constant. The average Fermi energies 
    $\epsilon_{{\rm F},i}$ of two reservoirs are the same, and the left reservoir is 
    set to be fully polarized, i.e., $h_{\rm L}/\epsilon_{\rm F}=1$ with $h_{\rm L}=\mu_{\up, {\rm L}}-\mu_{\dwn, {\rm L}}$. The temperature is set to be $T/T_{\rm F}=0.05$, which is close to zero.
    }
    \label{I1spin}
\end{figure}

\section{One-Body Spin-Tunneling Current}\label{One-body}

To study the spin-tunneling current between the reservoirs, we apply the Schwinger-Keldysh Green's function formalism~\cite{Schwinger,Keldysh}, which is adapted to nonequilibrium states with operators evolving with a bare Hamiltonian $H_0=\sum_{\bm{p},\sigma,i}(p^2/2m)c^\dagger_{\bm{p},\sigma,i}c_{\bm{p},\sigma,i}$.
On the other hand, we assume local equilibrium within each reservoir far from the junction, where operators in the interaction representation evolve with a grand-canonical Hamiltonian $K_0=H_0-\sum_{\bm{p},\sigma,i}\mu'_{\sigma,i}c^\dagger_{\bm{p},\sigma,i}c_{\bm{p},\sigma,i}$. The two reservoirs together with the junction constitute a nonequilibrium steady state. Notice that the relations between operators in the two representations read $c^{\dagger (H_0)}_{\bm{p},\sigma,i}(t)=e^{i\mu'_{\sigma,i}t}c_{\bm{p},\sigma,i}^{\dagger(K_0)}(t)$ and $c^{(H_0)}_{\bm{p},\sigma,i}(t)=e^{-i\mu'_{\sigma,i}t}c_{\bm{p},\sigma,i}^{(K_0)}(t)$. The perturbation theory gives the expression of spin current as 
\begin{align}
    I_{\rm s}(t,t')=&\sum_{n=0}^{\infty}\frac{(-i)^n}{n!}\int_Cdt_1\cdots\int_Cdt_n\nonumber\\
    &\langle {\rm T}_C \hat{I}_{\rm s}(t,t')H_{\rm T}(t_1)\cdots H_{\rm T}(t_n)\rangle.
\end{align}
The time integral in Eq.~(\ref{I1s1}) is taken along the Keldysh contour $C$, while ${\rm T}_C$ is the contour-time-ordering product. The different denotations $t$ and $t'$ are used to distinguish time arguments located on backward and forward branches on the Keldysh contour $C$.
Performing the expansion up to the leading order, the expectation value of the one-body spin-tunneling current is obtained as $I_{\rm 1s}\equiv\langle\hat{I}_{\rm 1s}(t,t)\rangle$, where
$\langle \cdots \rangle$ denotes the expectation value with respect to the nonequilibrium steady state and
\begin{align}\label{I1s1}
    \langle I_{\rm 1s}(t,t')\rangle=-\mathcal{T}^2_1\sum_{\bm{p},\bm{q},\sigma}&
    \int_Cdt_1\,\beta_{\sigma}\Big\langle{\rm T}_Ce^{i\mu'_{\sigma,{\rm R}}t}
    e^{-i\mu'_{\sigma,{\rm L}}t'}\nonumber\\
    &e^{i\Delta\mu'_{\sigma}t_1}c^\dagger_{\bm{q},\sigma,{\rm R}}(t)c_{\bm{p},\sigma,{\rm L}}(t')\nonumber\\
    &c^\dagger_{\bm{p},\sigma,{\rm L}}(t_1)c_{\bm{q},\sigma,{\rm R}}(t_1)\Big\rangle+{\rm H.c.}
\end{align}
with $\beta_{\up}=1$, $\beta_{\dwn}=-1$, and $\Delta\mu'_{\sigma}=\mu'_{\sigma,{\rm L}}-\mu'_{\sigma,{\rm R}}$. 
By using the Langreth rule and performing the Fourier transform, we write $I_{\rm 1s}$ as 
\begin{align}\label{I1s1_spectral}
    I_{\rm 1s}=4\mathcal{T}_1^2&\sum_{\bm{p},\bm{q},\sigma}\int\frac{d\omega}
    {2\pi}\beta_\sigma\big[\operatorname{Im}G^{\rm ret.}_{\bm{q},\sigma,{\rm L}}(\omega-\Delta\mu'_{\sigma})\nonumber\\
    &\times\operatorname{Im}G^{\rm ret.}_{\bm{p},\sigma,{\rm R}}(\omega)\big]\big[f(\omega-\Delta\mu'_\sigma)-f(\omega)\big],
\end{align}
where $G^{\rm ret.}$ is the retarded Green's function and $f(\omega)=1/(e^{\omega/T}+1)$ is the Fermi distribution function. These distribution functions are induced from the lesser propagators as $f(\omega)=-G^{<}(\omega)/[2i\operatorname{Im}G^{\rm ret.}(\omega)]$. We note that this one-body spin current is similar to the quasiparticle tunneling current~\cite{zhang2023dominant} except for the presence of the factor $\beta_\sigma$. Here, we adopt the zero-temperature Green's functions, $G^{\rm ret.}_{\bm{p},\sigma,i}(\omega)=1/(\omega-\xi^{\sigma}_{\bm{p},i}+i\eta)$, where $\eta$ is infinitesimal. The chemical potentials are calculated as $\mu_{\sigma,i}=dE/dn_{\sigma,i}=\epsilon_{{\rm F},\sigma,i}+gn_{\bar{\sigma},i}$,
where $\epsilon_{{\rm F},\sigma,i}$ is the Fermi energy of the spin-$\sigma$ component in the reservoir $i$. 
Although this expression of $\mu_{\sigma,i}$ is obtained at zero temperature, it can also be a reasonable value at low but nonzero temperatures.
We define the average Fermi energy $\epsilon_{{\rm F},i}=(\epsilon_{{\rm F},\up,i}+\epsilon_{{\rm F},\dwn,i})/2$ for each reservoir, and the average Fermi momentum $k_{F,i}$ is defined as $\epsilon_{{\rm F},i}=k^2_{{\rm F},i}/2m$. For the tunneling junction, we use a delta potential barrier $V(x)=V_0\delta(x/\lambda)$, yielding a constant $V(\bm{k})=V_0$ in the momentum space. Setting the average Fermi energy for both sides to be the same, i.e., $\epsilon_{\rm F,L}=\epsilon_{\rm F,R}=\epsilon_{\rm F}$, the averaged one-body tunneling amplitude $\mathcal{T}_1$ can be estimated as $\mathcal{T}_1=B_{\bm{0}}(\epsilon_{\rm F}+V_0)$, where $B_{\bm{0}}$ is the transmission coefficient for both components in the spin-balanced case~\cite{PhysRevA.106.033310,10.1093/pnasnexus/pgad045}. 

In Fig.~\ref{I1spin}, we show the one-body spin current $I_{1{\rm s}}\equiv \langle \hat{I}_{1{\rm s}}\rangle$.
Note that 
$X_1=9\pi\mathcal{T}_1^2\mathcal{N}^2/(4\epsilon_{\rm F})$ is a normalization constant with $\mathcal{N}=k^3_{\rm F}/3\pi^2$. 
The bias $\Delta h=h _ {\rm L} - h _ {\rm R}$ ranges between $0$ and $2\epsilon_{\rm F}$, where $\Delta h=2\epsilon_{\rm F}$ corresponds to the case that both reservoirs are fully-polarized but with opposite signs. 
At low temperatures, $I_{\rm 1s}$ exhibits an Ohmic transport.
This is similar to the quasiparticle tunneling through the junction with a chemical potential bias.
This trend can be found analytically by expanding the expression \eqref{I1s1_spectral} in powers of the chemical potential bias $\Delta \mu_{\sigma}'$.
Since the spectrum of the left reservoir is eventually independent of the bias,
$G^{\rm ret.}_{\bm{q},\sigma,{\rm L}}(\omega-\Delta\mu'_{\sigma}) = 1/(\omega-\xi^{\sigma}_{\bm{q},{\rm R}}+i\eta),$
the dependence stems from the distribution difference,
$f(\omega-\Delta\mu'_\sigma)-f(\omega) = (-\partial f/\partial \omega)\Delta\mu'_\sigma + \mathcal{O}({\Delta\mu'_\sigma} ^ 2).$
Thus, we can conclude the quasiparticle tunneling current is linearly dependent on the bias regardless of its carrier (i.e., spin or mass).

\section{Spin-Flip Tunneling Current}\label{spin-flip}

The spin-flip susceptibility, which can be used to characterize the ferromagnetic behavior of Fermi gases, plays an important role in investigating the dynamics of the spin-flip tunneling processes. 
In the spin-polarized gases, a dispersion of spin-flip collective modes (magnons) occurs in the spin-susceptibility spectra out from the Stoner particle-hole continuum~\cite{Sandri_2011}.
The dispersion facilitates the propagation of magnons. 

The linear response theory gives the spin-flip susceptibility as a retarded Green's function: 
\begin{equation}
    \chi^{\rm ret.}_{\bm{p},i}(t,t')=-i\theta(t-t')\left\langle S^+_{\bm{p},i}(t)S^-_{\bm{p},i}(t')+S^-_{\bm{p},i}(t')S^+_{\bm{p},i}(t)\right\rangle,
\end{equation}
where $S^+_{\bm{p},i}$ and $S^-_{\bm{p},i}$ are the spin ladder operators appearing in the spin-flip current operator in Eq.~(\ref{I2s}). 
By applying similar manipulations to those applied for $I_{\rm 1s}$, truncating the expression at the leading-order term, we can write 
\begin{align}\label{I2s1}
    I_{\rm 2s}=8\mathcal{T}_2^2\sum_{\bm{p},\bm{q}}\int\frac{d\omega}{2\pi}&\operatorname{Im}\chi^{\rm ret.}_{\bm{p},{\rm L}}(\omega) \operatorname{Im}\chi^{\rm ret.}_{\bm{q},{\rm R}}(\omega-2\Delta h)\nonumber\\
    &\times[b(\omega-2\Delta h)-b(\omega)],
\end{align}
where $b(\omega)=1/(e^{\omega/T}-1)$ is the Bose distribution function induced from $b(\omega)=\chi^<(\omega)/[2i\operatorname{Im}\chi^{\rm ret.}(\omega)]$. 
\begin{figure}[t]
    \centering
    \includegraphics[width=9cm]{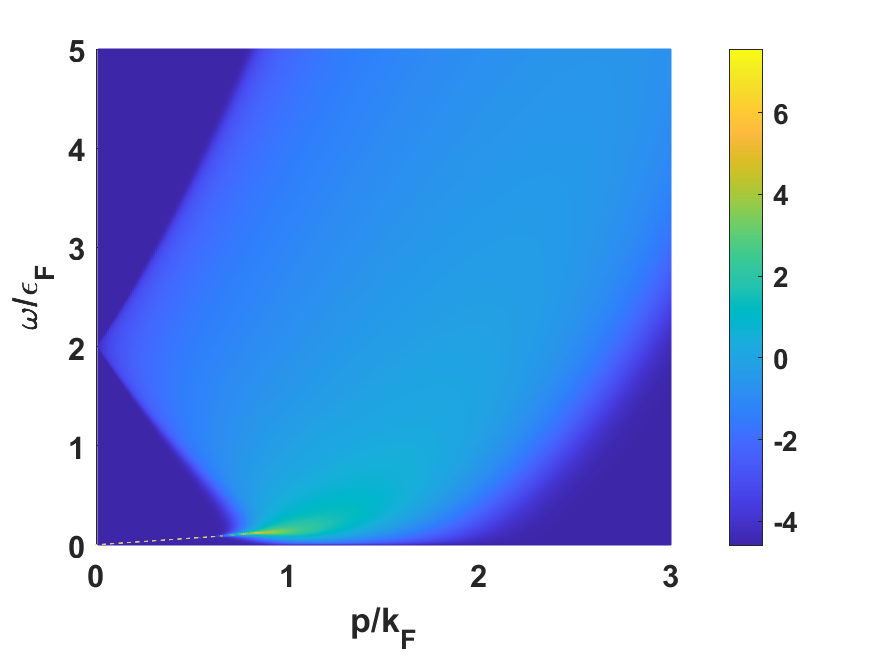}
    \caption{Imaginary part of the spin-flip susceptibility $\chi^{\rm ret.}_{\bm{p}}(\omega)$ for a fully-polarized two-component Fermi gas. The temperature is taken to be $T/T_{\rm F}=0.1$, and the two-body coupling strength is set to be $g^*=g\mathcal{N}/\epsilon_{\rm F}=\sqrt{2}$ to fulfill the gapless condition. The magnon dispersion can be seen at low momentum and low frequency and is indicated by the dotted eye guide. The color bar is shown with the logscale in arbitrary unit.}
    \label{ImX}
\end{figure}

In order to compute the spin-flip contribution $I_{\rm 2s}$, we utilize RPA~\cite{doi:10.1143/JPSJ.18.1025,ENGLERT1964429} to evaluate the spin-flip susceptibility. 
By employing energy representation, we can express
\begin{equation}
    \chi^{\rm ret.}_{\bm{p},i}(\omega)=\frac{\Pi_{\bm{p},i}(\omega)}
    {1+g\Pi_{\bm{p},i}(\omega)}.
\end{equation}
Here, $\Pi_{\bm{p},i}(\omega)$ is the Lindhard function,
\begin{equation}\label{Lindhard}
    \Pi_{\bm{p},i}(\omega)=\sum_{\bm{k}}\frac{f(\xi^+_{\bm{k},i})-
    f(\xi^-_{\bm{k+p},i})}{\omega-\xi^-_{\bm{k+p},i}+\xi^+_{\bm{k},i}+i\eta},
\end{equation}
where the symbol $+$ ($-$) denotes the spin with directions along (against) the polarization. According to RPA, the spin-susceptibility spectra, i.e., the imaginary part of $\chi_{\bm{p},i}$ can be obtained as 
\begin{equation}\label{RPA}
    \operatorname{Im}\chi^{\rm ret.}_{\bm{p},i}(\omega)=\frac{\operatorname{Im}
    \Pi_{\bm{p},i}(\omega)}{(1+g\operatorname{Re}\Pi_{\bm{p},i}(\omega))^2+
    (g\operatorname{Im}\Pi_{\bm{p},i}(\omega))^2}.
\end{equation}

\begin{figure}[t]
    \centering
    \subfigure{
    \includegraphics[width=7.8cm]{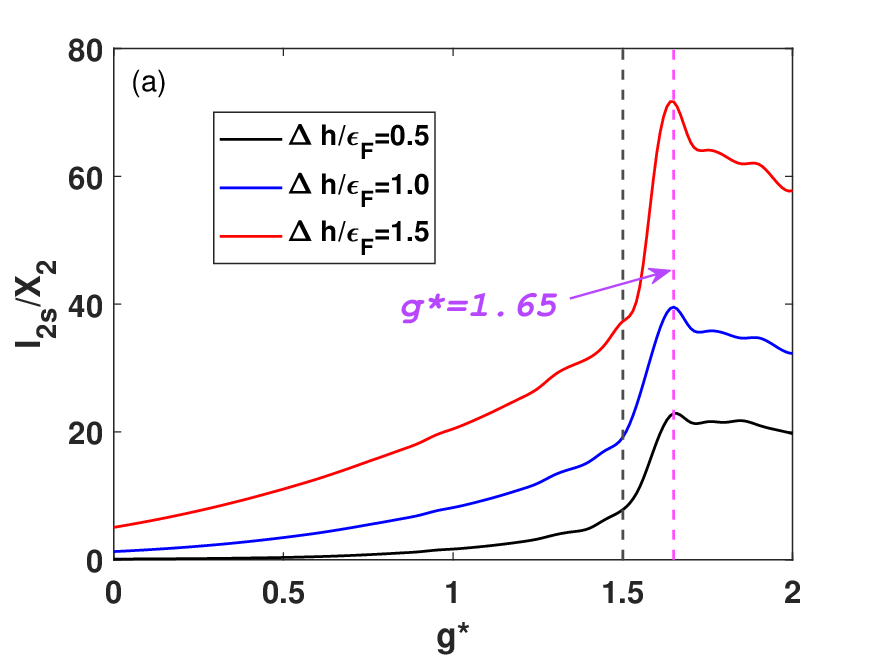}\label{I2s-G}
    }
    \subfigure{
    \includegraphics[width=7.8cm]{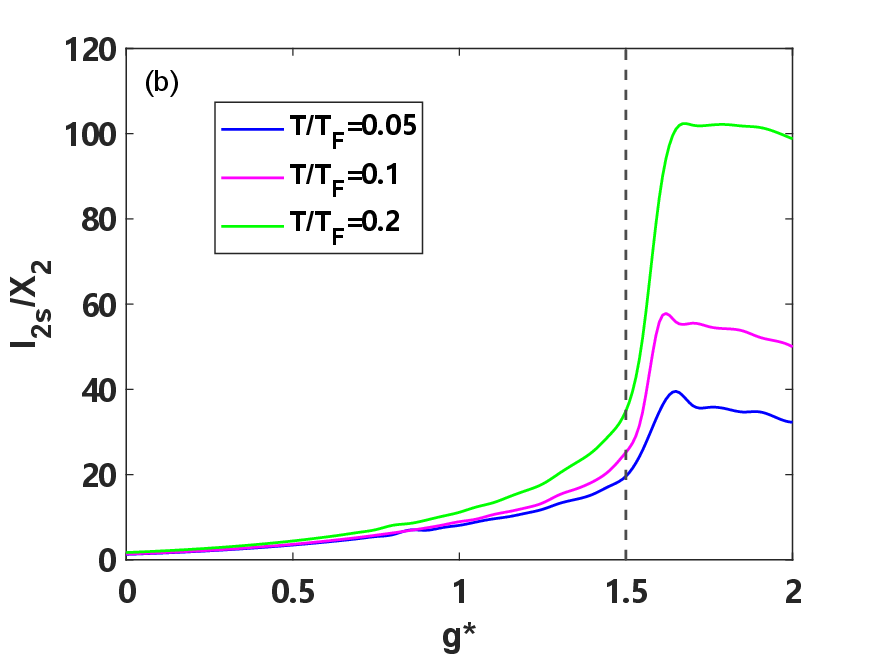}\label{I2s-G-T}
    }
    \caption{(a) The current-interaction features for $I_{\rm 2s}$ with different polarization bias $\Delta h$ at $T/T_{\rm F}=0.05$. (b) The interaction dependence of $I_{2s}$ with $\Delta h/\epsilon_F=1$ at different temperatures. The repulsive interaction strength is described by $g^*=8k_{\rm F}a/3\pi$. $X_2=9\mathcal{T}^2_2\mathcal{N}^4/(\pi\epsilon_{\rm F})$ is the normalizing constant. The current displays a sharp change at around $g^*=1.5$, which indicates a ferromagnetic phase transition.
    }
\end{figure}

Defining the normalized parameters $\tilde{k}=k/k_{\rm F}$, $\tilde{p}=p/k_{\rm F}$, and $\tilde{\omega}=\omega/\epsilon_{\rm F}$, we can write the real part of the Lindhard function as (see Appendix~\ref{appendixB})
\begin{align}\label{RePi}
    \operatorname{Re}\Pi_{\bm{p},i}(\omega)=\frac{3\mathcal{N}_i}{8\epsilon_{{\rm F},i}\tilde{p}}&\int d\tilde{k}\,\tilde{k}\Big\{ f(\xi^+_{\bm{k},i})\ln\big[A^+_{\tilde{\bm{p}},i}(\tilde{\omega},\tilde{k})\big]\nonumber\\
    &-f(\xi^-_{\bm{k},i})\ln\big[A^-_{\tilde{\bm{p}},i}(\tilde{\omega},\tilde{k})\big]\Big\},
\end{align}
where the amplitudes $A^{\pm}_{\tilde{\bm{p}},i}(\tilde{\omega},\tilde{k})$ are given by
\begin{equation}
    A^{\pm}_{\tilde{\bm{p}},i}(\tilde{\omega},\tilde{k})= \sqrt{\frac{[(\tilde{\omega}\mp\tilde{p}^2-2\tilde{h}_i)^2-4\tilde{k}^2\tilde{p}^2+\eta^2]^2+ (4\eta\tilde{k}\tilde{p})^2}{[(\tilde{\omega}-2\tilde{k}\tilde{p}\mp\tilde{p}^2-2\tilde{h}_i)^2+\eta^2]^2}}.
\end{equation}

\begin{figure}[t]
    \centering
    \includegraphics[width=8.6cm]{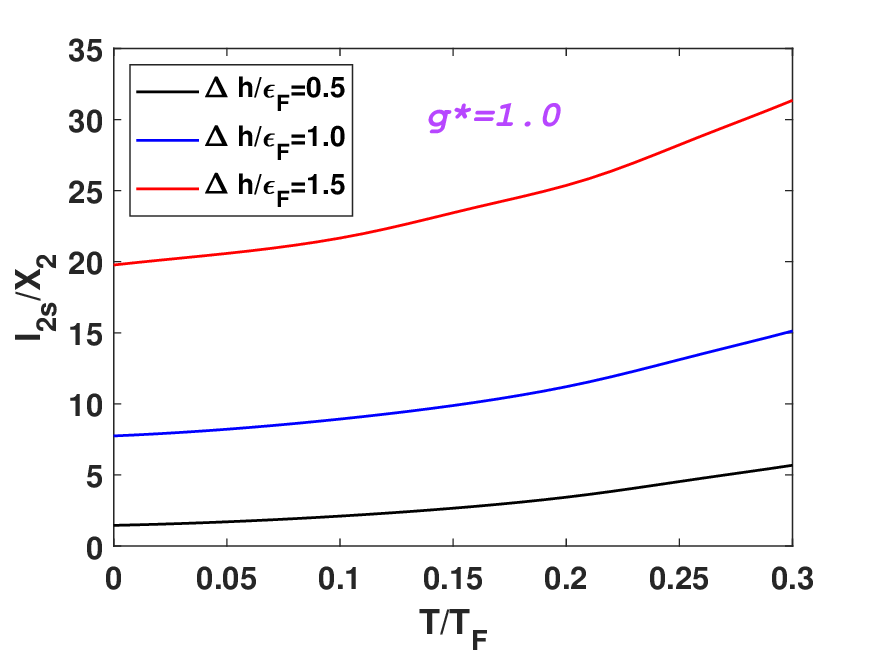}
    \caption{Temperature dependence of spin-flip tunneling current $I_{\rm 2s}$ with different values of $\Delta h$, where the coupling strength is set to be $g^*=1.0$. The results are obtained at temperatures that are relatively low compared with the Fermi temperature due to the zero-temperature approximation and chemical potential we applied.}
    \label{I2s-T}
\end{figure}

According to the Cauchy-Hadamard theorem, the imaginary part of the Lindhard function can be simply obtained as (see Appendix~\ref{appendixB})
\begin{equation}\label{ImPi}
    \operatorname{Im}\Pi_{\bm{p},i}(\omega)=-\frac{3\pi \mathcal{N}_i}{8\epsilon_{{\rm F},i}
    \tilde{p}}\int_\alpha^\infty d\tilde{k}\, \tilde{k}\big[f(\xi^+_{\bm{k},i})-f(\xi^-_{\bm{q}_0,i})\big],
\end{equation}
where we defined $\alpha=\frac{1}{2}\big|\frac{\tilde{\omega}-2\tilde{h}_i}{\tilde{p}}-\tilde{p}\big|$ and $\tilde{q}_0=\sqrt{\tilde{\omega}+\tilde{k}^2-2\tilde{h}_i}$.
Moreover, RPA develops a pole at $1+g\Pi_{\bm{p},i}(\omega)=0$, corresponding to the magnon peak.
When $\bm{p}\rightarrow 0$, the magnon pole appears at $\omega=2h-g(N_{\up}-N_{\dwn})$~\cite{doi:10.7566/JPSJ.90.024004}, which implies a possible energy gap for the magnon dispersion.
Defining $g^*=g\mathcal{N}/\epsilon_{\rm F}=8k_{\rm F}a/3\pi$, we can find that the zero-momentum pole for a fully polarized gas ($h/\epsilon_{\rm F}=1$) appears at $\omega=0$ when $g^*=\sqrt{2}$, which yields a gapless magnon dispersion.
Imposing such a gapless condition, the spin-susceptibility is shown in Fig.~\ref{ImX}. The dispersion of magnon modes, which manifests a quadratic law in $\bm{p}$ at small momentum~\cite{PhysRevB.32.2824}, can be seen as an apparent maximum below the Stoner particle-hole continuum.

\begin{figure}[t]
    \centering
    \subfigure{
    \includegraphics[width=7.8cm]{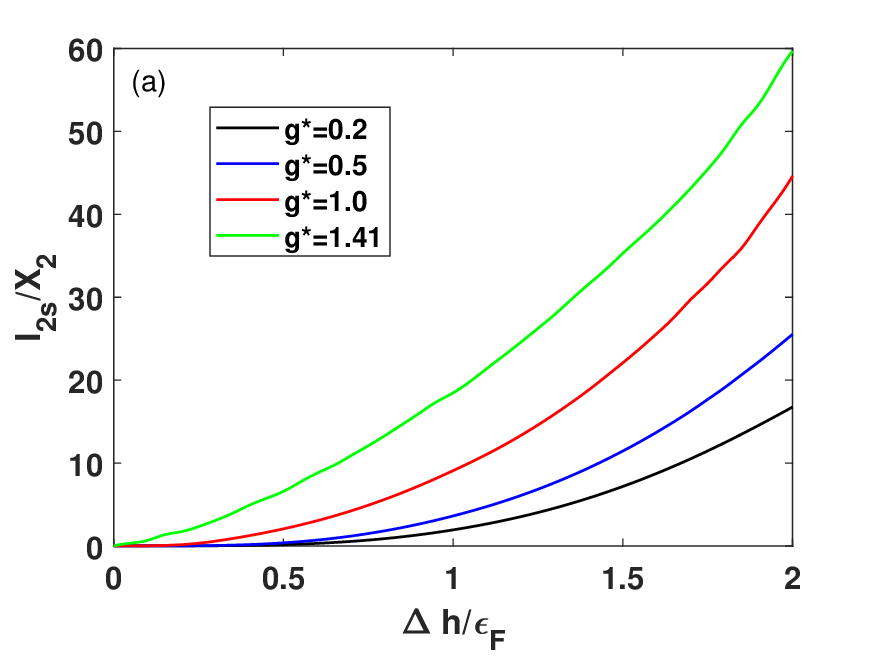}\label{I2spin}
    }
    \subfigure{
    \includegraphics[width=7.8cm]{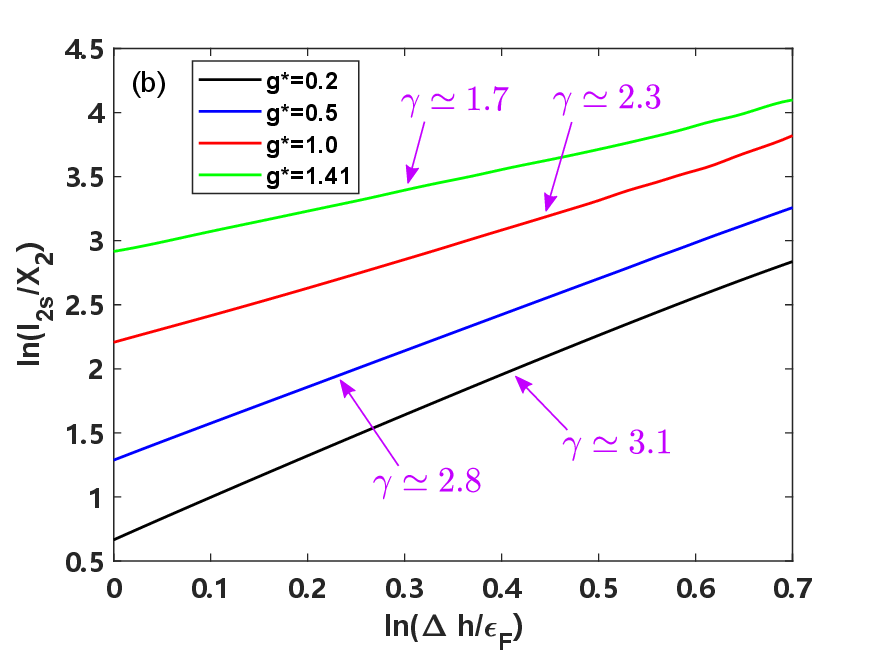}\label{log-log}
    }
    \caption{(a) The spin-flip tunneling current as a function of polarization bias $\Delta h$ with different two-body coupling strengths at $T/T_{\rm F}=0.05$. The average Fermi energies $\epsilon_{\rm F}$ and particle number density $\mathcal{N}$ of two reservoirs are set to be the same to induce a pure spin current. The left reservoir is set to be fully polarized, i.e., $h_{\rm L}/\epsilon_{\rm F}=1$, while the polarization of the right side is gradually tuned from $h_{\rm R}/\epsilon_{\rm F}=1$ to $h_{\rm R}/\epsilon_{\rm F}=-1$. (b) The log-log plot of the spin-flip current depending on the bias, where $\gamma$ represents the slope for each line.
    }
\end{figure}

Now, we are in a position to evaluate the spin-flip current within the leading order of $\mathcal{T}_2$ based on the spin-susceptibility spectra and Eq.~(\ref{I2s1}).
In the following, we are going to study the dependence of the current on the interaction strength, the system temperature, and the applied bias.

First, we investigate the interaction dependence of $I_{\rm 2s}$ in the polarized regime.
The Monte-Carlo calculations have predicted a critical interaction strength for an unpolarized repulsive Fermi gas as $k_{\rm F}a\simeq0.8$~\cite{PhysRevLett.103.207201,PhysRevLett.105.030405}, which corresponds to $g^*\simeq0.68$.
Figure~\ref{I2s-G} shows the current-interaction characteristics in the proposed polarized regime at $T/T_{\rm F}=0.05$. A critical value $g^*\simeq1.5$ ($k_{\rm F}a\simeq1.77$) is indicated for the ferromagnetic phase transition, as the current sharp increases and reaches a maximum at around $g^*=1.65$ ($k_{\rm F}a\simeq1.94$). This critical strength is close to the value at the gapless condition of magnons at zero temperature (i.e., $g^*=\sqrt{2}$), 
while the difference comes from the effect of the finite temperature as well as the finite practical value of $\eta$ in the numerical calculations. 
Also, it is worth noting that while this critical value is close to the Stoner's mean-field result, it is larger than the Monte Carlo result predicted from the extrapolation to non-polarized case~\cite{PhysRevLett.110.230401}. This difference may originate from the higher-ordered terms and the finite-range effect that are neglected in our analysis.
It is also observed that $I_{\rm 2s}$ gradually decreases above the critical repulsion strength.
This can be understood from the expression of spin susceptibility spectra in Eq.~\eqref{RPA}. At infinitely large interaction strength ($g^*\rightarrow\infty$), $\operatorname{Im}\chi^{\rm ret.}_{\bm{p},i}(\omega)$ tends to vanish, indicating the Stoner continuum and magnon excitation are suppressed.
More intuitively, if the repulsive interaction is strong, the left reservoir cannot further transfer the spin $\sigma=\uparrow$ to the right reservoir because of the strong repulsive interaction with the $\sigma=\downarrow$, leading to the suppression of $I_{\rm 2s}$. 
As a result,
$I_{\rm 2s}$
approaches zero in the strong-repulsion limit. However, one should notice that over the critical interaction strength ($g^*=\sqrt{2}$ at $T=0$, which becomes larger at $T\neq 0$), the system becomes metastable via the first-order transition towards the inhomogeneous phase when going beyond the mean-field theory~\cite{science.1177112,PhysRevLett.104.220403,PhysRevLett.105.030405}. 
In this regard, our results in the strong-repulsion regime may be regarded as the spin-flip tunneling transport under the metastable condition.
As experimentally reported in Ref.~\cite{PhysRevLett.129.203402}, the upper bound of the coupling strength for equilibrium repulsive Fermi gases is $k_{\rm F}a\simeq 1$. Beyond this coupling, we need to consider the three-body loss effect, which is out of the scope in this paper.

Meanwhile, the effect of temperature on the current-interaction characteristics can be seen in Fig.~\ref{I2s-G-T}, where all temperatures are set to be relatively low as we use the low-temperature approximation
(i.e., we adopted the zero-temperature propagators and included the temperature dependence in the distribution functions).
We can see that no significant shift of the critical point has been observed apparently. 
To obtain the results more precisely, other finite-temperature corrections such as damping of quasiparticles need to be considered, which is left for future work.
Here, we incorporate the temperature variations of the distribution functions to gain a preliminary understanding of the temperature dependency of the currents.
Figure~\ref{I2s-T} depicts the temperature dependence of $I_{\rm 2s}$ with various values of $\Delta h$.
It is worth noting that both one-body and two-body spin tunnelings are enhanced at high temperatures.
This is a consequence of the properties of the Fermi and Bose distributions.
Note that we have chosen the coupling strength to be $g^*=1.0$ as a representative, as other coupling strengths demonstrate similar temperature dependence.

To obtain the current-bias feature, the left reservoir is set to be always fully polarized ($h_{\rm L}/\epsilon_{\rm F}=1$), while the polarization of the right side varies between $h_{\rm R}/\epsilon_{\rm F}=1$ and $h_{\rm R}/\epsilon_{\rm F}=-1$.
Here we focus on the weakly-repulsive regime ($g^*\leq \sqrt{2}$), where the homogeneous right reservoir is stable against the phase separation (i.e., ferromagnetism).
Notice that the spin-flip current $I_{\rm 2s}$ for various coupling strength $g ^ *$ becomes larger as the bias $\Delta h$ increases [Fig.~\ref{I2spin}]. 
On the other hand, according to Eq.~(\ref{RPA}), the Stoner continuum is suppressed by the large $g^*$, which indicates that the transport of magnon modes plays a major role in strong coupling regimes.
Moreover, the spin-flip current exhibits a nonlinear dependence on $\Delta h$, which is different from the behavior of $I_{\rm 1s}$.
We suppose that the current $I_{\rm 2s}$ is a power function of the bias $\Delta h$ like $I_{\rm 2s}\propto\Delta h^\gamma$ while the log-log plot shown in Fig.~\ref{log-log} numerically gives the value of $\gamma$.
The exponent is obtained as $\gamma\simeq 3.1$ at $g^*=0.2$ and decreases when the interaction becomes stronger. 
This indicates the current is more sensitive to the changes of the bias $\Delta h$ in the weakly interacting case.
If we expand the spectrum $\operatorname{Im}\chi^{\rm ret.}_{\bm{p},{\rm L}}(\omega-2\Delta h)$ and the distribution function $b(\omega-2\Delta h)$ in Eq.~(\ref{I2s1}) in terms of the bias $\Delta h$ and keep to the third order, we will find that the second-order term vanishes while the first- and third-order terms remain, leading to an odd function which is consistent with the anti-symmetry of current with respect to the bias.
The sensitivity of current to the bias may be caused by the dominance of the third-order term in the weak coupling side.
Such a different dependence with respect to the polarization bias may provide a way to distinguish the one-body and spin-flip (two-body) tunneling signals.
More interestingly, the significance of nonlinear dependence in the weak coupling side may provide a possible way to induce a third harmonic spin current by applying an AC spin bias on the junction,
which enables us to clearly distinguish the two-body signal from the one-body signal in the frequency domain. 
Note that the oscillation period of AC bias should be comparable to the timescale of the tunneling process, which can be estimated by the uncertainty principle, so that the present results can be applied adiabatically.

\section{Conclusion}\label{conclusion}

In this study, we have theoretically studied the spin tunneling current induced by a magnetization bias between two repulsively interacting Fermi gases near the ferromagnetic phase transition.
Utilizing the Schwinger-Keldysh formalism, we have derived the one-body and spin-flip tunneling currents up to the leading-order of the single-particle wave-function amplitude near the potential barrier.
Based on the spin-flip susceptibility functions with RPA, we have computed the spin-flip current. 
We have shown how the one-body spin current and spin-flip current vary with the polarization bias between two gases. 
The one-body contribution increases linearly with the bias, while the spin-flip one exhibits a predominantly cubic dependence.
This nonlinearity implies the generation of third harmonics in the spin current when an AC bias is applied.
We have also investigated the interaction and temperature dependencies of the spin-flip current in the present system.
For fully polarized Fermi gases, a critical repulsive strength is demonstrated close to the gapless conditions.
The magnon modes, which appear as poles in the spin-susceptibility spectra, are supposed to play a major role in the spin tunneling processes in the strong-coupling regime (large $g$).
Moreover, our study may provide a practical tool for estimating the coupling strengths of one-body and spin-flip tunnelings in cold atomic systems.

\begin{acknowledgements}
T.Z. thanks Z. Lyu and T. Chen for the technical help.
T.Z.~is supported by the RIKEN Junior Research Associate Program.
D.O.~is supported by the JSPS Overseas Research Fellowship, by the Institution of Engineering and Technology (IET), and by Funda\c{c}\~ao para a Ci\^encia e a Tecnologia and Instituto de Telecomunica\c{c}\~oes under project UIDB/50008/2020.
H.T.~is supported by the JSPS KAKENHI under Grants Nos.~18H05406, 22H01158, and 22K13981. M.M.~is supported by the JSPS KAKENHI under Grants Nos.~21H01800, 21H04565, and 23H01839.
H.L.~is supported by the JSPS KAKENHI under Grant No.~20H05648 and the RIKEN Pioneering Project: Evolution of Matter in the Universe.
The authors thank RIKEN iTHEMS NEW working group for fruitful discussions.
\end{acknowledgements}

\appendix

\section{Derivation of Hamiltonian}\label{appendixA}

The Hamiltonian for the two-terminal model connected through a potential barrier with a contact-type two-body interaction in each bulk system is given by
\begin{align}\label{H}
    \hat{H}=&\int d^3\bm{r}\sum_{\sigma}\hat{\psi}^\dagger_\sigma(\bm{r})\Big(-\frac{\nabla^2}{2m}+V(\bm{r})\Big)\hat{\psi}_\sigma(\bm{r})\nonumber\\
    &+g\int d^3\bm{r}\hat{\psi}^\dagger_{\uparrow}(\bm{r})\hat{\psi}^\dagger_{\downarrow}(\bm{r})\hat{\psi}_{\downarrow}(\bm{r})\hat{\psi}_{\uparrow}(\bm{r}),
\end{align}
where $\hat{\psi}_{\sigma}(\bm{r})$ denotes the field operator for wave functions of particles with spin $\sigma$, $V(\bm{r})$ describes the potential barrier, and $g=4\pi a/m$ is the two-body coupling constant with the $s$-wave scattering length $a$. Notice that the field operator $\hat{\psi}_{\sigma}(\bm{r})$ can be rewritten as $\hat{\psi}_{\sigma}(\bm{r})=\hat{\psi}_{\sigma,{\rm L}}(\bm{r})+\hat{\psi}_{\sigma,{\rm R}}(\bm{r})$. Inserting it into the Hamiltonian above, we have the local reservoir Hamiltonian: 
\begin{align}\label{Hi}
    \hat{H}_{i={\rm L,R}}&=\int d^3\bm{r}\sum_{\sigma}\hat{\psi}^\dagger_{\sigma,i}(\bm{r})\Big(-\frac{\nabla^2}{2m}\Big)\hat{\psi}_{\sigma,i}(\bm{r})\nonumber\\
    &+g\int d^3\bm{r}\hat{\psi}^\dagger_{\uparrow,i}(\bm{r})\hat{\psi}^\dagger_{\downarrow,i}(\bm{r})\hat{\psi}_{\downarrow,i}(\bm{r})\hat{\psi}_{\uparrow,i}(\bm{r}),
\end{align}
and the one-body tunneling term:
\begin{align}\label{H1T}
    \hat{H}_{\rm 1T}&=\int d^3\bm{r} \sum_{\sigma}\Big[\hat{\psi}^\dagger_{\sigma,{\rm L}}(\bm{r})\Big(-\frac{\nabla^2}{2m}\nonumber\\
    &+g\sum_{i}\hat{N}_{\bar{\sigma},i}(\bm{r})\Big)
    \hat{\psi}_{\sigma,{\rm R}}(\bm{r})+{\rm H.c.}\Big],
\end{align}
where $\hat{N}_{\sigma,i}(\bm{r})$ is the density operator. Also, we can obtain the pair tunneling term: 
\begin{equation}\label{Hpair}
     \hat{H}_{\rm pair}=g \int d^3 \boldsymbol{r}\left[\hat{P}_{\mathrm{L}}^{\dagger}(\boldsymbol{r}) \hat{P}_{\mathrm{R}}(\boldsymbol{r})+\text { H.c. }\right] 
\end{equation}
where $\hat{P}^\dagger_i(\bm{r})=\hat{\psi}^\dagger_{\uparrow,i}(\bm{r})\hat{\psi}^\dagger_{\downarrow,i}(\bm{r})$ is the pair creation operator, and the spin-flip tunneling term:
\begin{equation}\label{H2T}
    \hat{H}_{2T}=g \int d^3 \bm{r}\left[\hat{S}_{\mathrm{L}}^{+}(\bm{r}) \hat{S}_{\mathrm{R}}^{-}(\bm{r})+\hat{S}_{\mathrm{R}}^{+}(\bm{r}) \hat{S}_{\mathrm{L}}^{-}(\bm{r})\right] 
\end{equation}
with the spin ladder operators $\hat{S}_i^{+}(\bm{r})=\hat{\psi}_{\uparrow, i}^{\dagger}(\bm{r}) \hat{\psi}_{\downarrow, i}(\bm{r})$ and $\hat{S}_i^{-}(\bm{r})=\hat{\psi}_{\downarrow, i}^{\dagger}(\bm{r}) \hat{\psi}_{\uparrow, i}(\bm{r})$. Notice that we can omit the pair-tunneling coupling since the pair-tunneling current does not occur because we consider the vanishing chemical-potential bias ($\mu_{\rm L}-\mu_{\rm R}=0$).

While the potential barrier peaking in the junction between the reservoirs may induce an inhomogeneity near the barrier, far from the junction the potential goes smoothly to zero. Therefore, we can consider uniform gases inside the reservoirs, with the wave function being the asymptotic form: 
\begin{equation}
\label{eq:psi_l}
    \psi_{\sigma, \mathrm{L}}(\boldsymbol{r})=\sum_{\boldsymbol{p}} \widetilde{c}_{\boldsymbol{p}, \sigma, \mathrm{L}}\times \begin{cases}e^{i \boldsymbol{p} \cdot \boldsymbol{r}}+R_{\boldsymbol{p}, \sigma} e^{-i \boldsymbol{p} \cdot \boldsymbol{r}} &(x<0), \\ B_{\bm{p}, \sigma} e^{i \boldsymbol{p} \cdot \boldsymbol{r}} &(x>0),\end{cases}
\end{equation}
\begin{equation}
\label{eq:psi_r}
    \psi_{\sigma, \mathrm{R}}(\boldsymbol{r})=\sum_{\bm{p}} \widetilde{c}_{\boldsymbol{p}, \sigma, \mathrm{R}}\times \begin{cases}B_{\bm{p}, \sigma} e^{-i \boldsymbol{p} \cdot \boldsymbol{r}} &(x<0), \\ e^{-i \bm{p} \cdot \boldsymbol{r}}+R_{\bm{p}, \sigma} e^{i \bm{p} \cdot \boldsymbol{r}}& (x>0),\end{cases}
\end{equation}
where $\widetilde{c}_{\bm{p},\sigma,i}$ is the amplitude of the asymptotic wave function while $R_{\bm{p},\sigma}$ and $B_{\bm{p},\sigma}$ are respectively one-particle reflection and transmission coefficients with respect to the potential barrier.
In Eqs.~\eqref{eq:psi_l} and \eqref{eq:psi_r},
$x$ symbolically denotes the direction perpendicular to the potential barrier at $x=0$.
By substituting the asymptotic wave functions into Eq.~(\ref{Hi})-(\ref{H2T}), and replacing $\widetilde{c}_{\bm{p},\sigma,i}$ with the fermionic annihilation operator $c_{\bm{p},\sigma,i}$, we obtain the reservoir Hamiltonian and tunneling Hamiltonian as Eq.~(\ref{Hreservoir}) and Eq.~(\ref{Htunneling}).

\section{Calculation of Spin-Flip Susceptibility}\label{appendixB}

In this appendix, we give the details of calculations of the spin-flip susceptibility spectra. We notice that Eq.~(\ref{Lindhard}) can be rewritten as 
\begin{align}
    \Pi_{\bm{p},i}(\omega)=&\sum_{\bm{k}}\frac{f(\xi^+_{\bm{k},i})}
    {\omega-\xi^-_{\bm{k+p},i}+\xi^+_{\bm{k},i}+i\eta}\nonumber\\
    &-\sum_{\bm{k}}\frac{f(\xi^-_{\bm{k},i})}
    {\omega-\xi^-_{\bm{k},i}+\xi^+_{\bm{k-p},i}+i\eta}.
\end{align}
By changing the discrete summation over $\bm{k}$ into the integral over parameters in a spherical coordinate and carrying out the angular integral, we obtain 
\begin{align}
        \Pi_{\bm{p},i}(\omega)&=\frac{k_{{\rm F},i}^3}{8\pi^2\epsilon_{{\rm F},i}\tilde{p}}\int d\tilde{k}\,\tilde{k}\nonumber\\ &\Bigg[f(\xi^+_{\bm{k},i})\ln\Bigg(\frac{\tilde{\omega}+i\eta+2\tilde{k}\tilde{p}-\tilde{p}^2-2\tilde{h}_i}{\tilde{\omega}+i\eta-2\tilde{k}\tilde{p}-\tilde{p}^2-2\tilde{h}_i}\Bigg)\nonumber\\
        &-f(\xi^-_{\bm{k},i})\ln\Bigg(\frac{\tilde{\omega}
        +i\eta+2\tilde{k}\tilde{p}+\tilde{p}^2-2\tilde{h}_i}{\tilde{\omega}+i\eta-2\tilde{k}
        \tilde{p}+\tilde{p}^2-2\tilde{h}_i}\Bigg)\Bigg].
\end{align}
Due to the infinitesimally small number $\eta$, the formulas in the parentheses can be expressed as 
\begin{widetext}
\begin{align}
    \frac{\tilde{\omega}+i\eta+2\tilde{k}\tilde{p}\mp\tilde{p}^2-2\tilde{h}_i}
    {\tilde{\omega}+i\eta-2\tilde{k}\tilde{p}\mp\tilde{p}^2-2\tilde{h}_i}=&
    \sqrt{\frac{[(\tilde{\omega}\mp\tilde{p}^2-2\tilde{h}_i)^2-4\tilde{k}^2\tilde{p}^2+\eta^2]^2+16\eta^2\tilde{k}^2\tilde{p}^2}{[(\tilde{\omega}-2\tilde{k}\tilde{p}
    \mp\tilde{p}^2-2\tilde{h}_i)^2+\eta^2]^2}}\nonumber\\
    &\times\exp\Bigg\{-i\arctan\Bigg[\frac{4\tilde{k}\tilde{p}\eta}{(\tilde{\omega}\mp\tilde{p}^2-2\tilde{h}_i)^2-4\tilde{k}^2\tilde{p}^2+\eta^2}\Bigg]\Bigg\}
    \equiv A^{\pm}_{\tilde{\bm{p}},i}(\tilde{\omega},\tilde{k})e^{-i\theta^{\pm}_{\bm{p},i}(\tilde{\omega},\tilde{k})}.
\end{align}
\end{widetext}
Thus by defining $\mathcal{N}_i=k^3_{{\rm F},i}/3\pi^2$, we obtain the real part of Lindhard function expressed as Eq.~(\ref{RePi}). 

On the other hand, according to Cauchy-Hadamard principal value theorem: $\frac{1}{\Gamma+i\eta}=\mathcal{P}\frac{1}{\Gamma}-i\pi\delta(\Gamma)$, we may write the imaginary part of Lindhard function as 
\begin{align}\label{ImPi1}
    \operatorname{Im}\Pi_{\bm{p},i}(\omega)=&-\pi\int\frac{d^3\bm{k}}{(2\pi)^3}\big[f(\xi^+_{\bm{k},i}-f(\xi^-_{\bm{k}+\bm{p},i})\big]\nonumber\\
    &\times\delta\left[\omega-\frac{(\bm{k}+\bm{p})^2}{2m}+\frac{\bm{k}^2}{2m}-2h_i\right].
\end{align}
By writing the integral over parameters in a spherical coordinate and performing the change of variables, $\cos{\theta}\rightarrow q=|\bm{k}+\bm{p}|$, Eq.~(\ref{ImPi1}) can be rewritten as 
\begin{align}
    \operatorname{Im}\Pi_{\bm{p},i}(\omega)=&-\frac{k^{3}_{{\rm F},i}}{4\pi\epsilon_{{\rm F},i}\tilde{p}}\int d\tilde{k}\, \tilde{k}\int_{|\tilde{k}-\tilde{p}|}^{\tilde{k}+\tilde{p}} d\tilde{q}\, \tilde{q}\big[f(\xi^+_{\bm{k},i})\nonumber\\
    &\quad -f(\xi^-_{\bm{k}+\bm{p},i})\big]\delta(\tilde{\omega}-\tilde{q}^2+\tilde{k}^2-2\tilde{h}_i),
\end{align}
where $\tilde{q}=q/k_{\rm F}$. Then by using the identity $\delta[f(x)]=\delta(x-x_0)/|f'(x_0)|$ with $f(x_0)=0$, we have
\begin{align}
    \operatorname{Im}\Pi_{\bm{p},i}(\omega)=&-\frac{k^{3}_{{\rm F},i}}{4\pi\epsilon_{{\rm F},i}\tilde{p}}\int d\tilde{k}\, \tilde{k}\int_{|\tilde{k}-\tilde{p}|}^{\tilde{k}+\tilde{p}} d\tilde{q}\, \tilde{q}\nonumber\\
    &\quad \times \big[f(\xi^+_{\bm{k},i})-f(\xi^-_{\bm{k}+\bm{p},i})\big]\delta(\tilde{q}-\tilde{q}_0),
\end{align}
where $\tilde{q}_0=\sqrt{\tilde{\omega}+\tilde{k}^2-2\tilde{h}_i}$. Meanwhile, to make the integral be nonzero, $\tilde{q}_0$ should satisfy the inequality $|\tilde{k}-\tilde{p}|\leq \tilde{q}_0\leq \tilde{k}+\tilde{p}$, which yields a lower limit of integral over $k$ as
\begin{equation}
    k\geq\frac{1}{2}\bigg|\frac{\tilde{\omega}-2\tilde{h}_i}{\tilde{p}}-\tilde{p}\bigg|.
\end{equation}
Therefore, we gain the expression of the imaginary part of Lindhard as Eq.~(\ref{ImPi}). By defining $g^*=g\mathcal{N}/\epsilon_{\rm F}$, we are able to calculate the imaginary part of $\chi_{\bm{p},i}(\omega)$ as 
\begin{equation}
    \operatorname{Im}\tilde{\chi}_{\bm{p},i}(\omega)=
    \frac{\operatorname{Im}\tilde{\Pi}_{\bm{p},i}(\omega)}{(1+g^*\operatorname{Re}
    \tilde{\Pi}_{\bm{p},i}(\omega))^2+(g^*\operatorname{Im}\tilde{\Pi}_{\bm{p},i}
    (\omega))^2},
\end{equation}
where $\tilde{\chi}_{\bm{p},i}=\chi_{\bm{p},i}\epsilon_{{\rm F},i}/\mathcal{N}_i$ and $\tilde{\Pi}_{\bm{p},i}=\Pi_{\bm{p},i}\epsilon_{{\rm F},i}/\mathcal{N}_i$.

\bibliographystyle{apsrev4-1}
\bibliography{ref.bib}

\end{document}